\documentclass[sigconf, nonacm]{acmart}
\AtBeginDocument{%
  }


\newif\iffinal
\iffinal
    \newcommand\joaquin[1]{}
    \newcommand\robert[1]{}
    \newcommand\chandra[1]{}
    \newcommand\igor[1]{}
    \newcommand\raj[1]{}
    \newcommand\dario[1]{}
    {}
\else
\usepackage{xcolor}
    \newcommand\joaquin[1]{{\color{orange}[Joaquin: #1]}}
    \newcommand\robert[1]{{\color{red}[Robert: #1]}}
    \newcommand\chandra[1]{{\color{cyan}[Chandra: #1]}}
    \newcommand\igor[1]{{\color{purple}[Igor: #1]}}
    \newcommand\raj[1]{{\color{green}[Raj: #1]}}
    
\fi

\usepackage{svg}
\usepackage{siunitx}
\usepackage{tikz}
\usetikzlibrary{positioning,arrows.meta, decorations.pathreplacing}
\usepackage{comment}
\usepackage{subcaption}
\usepackage{caption}
\usepackage{multirow}
\usepackage{array}
\usepackage{booktabs}
\usepackage{svg}
\usepackage{hyperref}

\begin{document}

\title{Beyond Assumptions: Measuring Federated Learning over Real 5G Networks}

\author{Robert J. Hayek}
\affiliation{%
 \department{Data Science and Learning}
 \institution{Argonne National Laboratory}
 \state{Illinois}
 \country{USA}}
\email{rhayek@anl.gov}
\authornote{These authors contributed equally to this work.}
\orcid{0000-0002-4015-5897}

\author{Kayla Comer}
\affiliation{%
  \department{Electrical and Computer Eng.}
  \institution{Northwestern University}
  \state{Illinois}
  \country{USA}}
\email{kcomer@u.northwestern.edu}
\authornotemark[1]
\orcid{0000-0002-9705-7243}

\author{Joaquin Chung}
\affiliation{%
  \department{Data Science and Learning}
  \institution{Argonne National Laboratory}
  \state{Illinois}
  \country{USA}}
\email{chungmiranda@anl.gov}
\orcid{0000-0001-7383-3810}

\author{Chandra R. Murthy}
\affiliation{%
  \department{Electrical Communication Eng.}
  \institution{Indian Institute of Science}
  \state{Bangalore}
  \country{India}}
\email{cmurthy@iisc.ac.in}
\authornote{C.R.~Murthy's initial contribution to this work was performed while he was at Argonne National Laboratory and Northwestern University.}
\orcid{0000-0003-4901-9434}

\author{Rajkumar Kettimuthu}
\affiliation{%
  \department{Data Science and Learning}
  \institution{Argonne National Laboratory}
  \state{Illinois}
  \country{USA}}
\email{kettimut@anl.gov}
\orcid{0000-0002-0046-9883}

\author{Igor Kadota}
\affiliation{%
  \department{Electrical and Computer Eng.}
  \institution{Northwestern University}
  \state{Illinois}
  \country{USA}}
\email{kadota@northwestern.edu}
\orcid{0000-0002-9075-3377}

\renewcommand{\shortauthors}{R. J. Hayek, K. Comer et al.}

\begin{abstract}
Deploying FL using IoT devices is an area poised to significantly benefit from advances in NextG wireless. In this paper, we deploy a FL application using a 5G-NR Standalone (SA) testbed with open-source and Commercial Off-the-Shelf (COTS) components. The 5G testbed architecture consists of a network of resource-constrained edge devices, namely Raspberry Pis, and a central server equipped with a Software Defined Radio (SDR) and running O-RAN software. Our testbed allows edge devices to communicate with the server using WiFi and Ethernet in addition to 5G. FL is deployed using the Flower FL framework, extended with custom instrumentation for communication and ML metrics. We analyze the FL application across three network interfaces--5G, WiFi, and Ethernet--as well as across 5G bandwidths and uplink-downlink scheduling ratios. Our experimental results challenge some common assumptions about communication time in FL over wireless and discuss the potential pitfalls of these assumptions. We find that there is a consistent straggler in about 70\% of trials, while in the other 30\%, high communication time causes competing stragglers. We also compare FL performance over 5G with and without external congestion and compare our testbed to commercial 5G to validate our findings in a broader context. For reproducibility, we have open-sourced our FL application, instrumentation tools, and testbed configuration.
\end{abstract}


\maketitle

\section{Introduction} \label{sec:intro}
Federated Learning (FL) is a method of distributed machine learning that utilizes a collection of decentralized datasets across a communication network to train a global model. Since the introduction of the first algorithm, Federated Averaging (FedAvg), FL has emerged as the primary method of privacy-aware and communication-efficient collaborative learning, as it does not require the transfer of raw data but model parameters~\cite{mcmahan_communication-efficient_2017, banabilah_federated_2022}. However, the main challenge of FL is the constraint imposed by the communication network due to the frequent exchange of model parameters. Though more efficient than transferring data to one location for centralized training, FL still requires gigabytes of data transfer for simple image classification tasks~\cite{decentralized-sun,ma_adaptive_2023}. However, despite the recognized importance of communication networks, only a few papers have implemented and evaluated FL over real wireless networks.

A large body of work exists regarding the analysis and optimization of FL architectures over communication networks (see surveys in~\cite{banabilah_federated_2022, lim_federated_2020}). We discuss a few representative studies and highlight works that are most related to this paper. Several papers have sought to reduce the communication overhead through improved client selection and scheduling methods~\cite{skocaj_uplink_2023, zang_general_2023, Alanazi-fedband, pase-imperfect-csi, narmadha_fedeff_2025, wang-fog, chen-multihop, chen_convergence_2021}, decentralized or asynchronous models~\cite{decentralized-sun, yin_decentralized_2023, Zehtabi-decentralized-sporadic, zhang-automotive, sad-async-peertopeer, fraboni_general}, over-the-air aggregation~\cite{zhang2024uplinkovertheairaggregationmultimodel, wang-adaptive-ota}, adaptive models~\cite{ma_adaptive_2023, deng-sparse-adaptive, jiang_model_2023}, and split learning~\cite{mu-split}. Recent work on FL over wireless networks can be classified into three categories: theory/simulation-based studies, emulated network implementations, and real-world deployments. Each provides valuable information about the performance and limitations of FL.

Simulation-based/theoretical studies provide insights into the expected behavior of an FL application over wireless and propose optimizations backed by simulation results. In this area of work, authors make several assumptions related to channel quality, compute time, and communication time---either implicitly or explicitly. Simulation or theoretical studies are helpful in understanding the potential optimizations and shortcomings of FL over wireless. Researchers have identified unique limitations that must be addressed for wireless deployment, such as power-limited devices, wireless channel properties, limited available bandwidth, and security issues, as discussed in the recent survey~\cite{beitollahi_federated_2023}. In~\cite{chen_convergence_2021}, the authors assert that a fundamental bottleneck of FL over wireless systems is its limited communication capacity, which limits model complexity and node participation. Additionally, the authors propose a theoretical framework backed by simulation results that uses a probabilistic user selection scheme to reduce convergence time by $56\%$. Other works have focused on optimizing resource allocation~\cite{tran_federated_2019}, addressing the effects of stragglers over a wireless network~\cite{zang_general_2023}, and optimizing device scheduling to reduce convergence time~\cite{wan_convergence_2021}. Alternatives to traditional FedAvg have also been proposed. These collaborative FL frameworks aim to improve upon convergence time while requiring less reliance on central controllers~\cite{chen_wireless_2020}. The authors propose techniques to optimize collaborative FL deployments in IoT systems to address issues related to the loss function, convergence time, energy consumption, and link reliability. 

To close the gap between simulation and implementation, FL is implemented in emulated network environments, which typically involves the full network stack, but lacks the physical RAN and/or the physical wireless channel. For example, testbeds like Colosseum provide platforms for large-scale experimental research on emulated RAN with real software-defined radios but emulated wireless channels~\cite{bonati_colosseum_2021}. 
In~\cite{rumesh_federated_2024}, the authors utilized this testbed to demonstrate a federated learning system over an ORAN emulator. Additionally, in~\cite{nardini_scalable_2021}, the authors develop a 5G testbed capable of supporting hundreds of emulated UEs.

In the context of 5G networks, OpenAirInterface (OAI) has emerged as a promising platform for experimental research~\cite{nikaein_openairinterface_2014}. In~\cite{Lee_Shin_2022}, the authors implemented a functioning 5G testbed using open-source components to evaluate FL. They utilize the open-source free5GC~\cite{free5gc} core network and use a simulated RAN, to implement an FL application inside a distributed NWDAF architecture. Similarly, in~\cite{Rajabzadeh_Outtagarts_2023}, the authors propose an FL strategy leveraging virtual NWDAF instances to implement a modified approach using influence-based weighted FedAvg without simulation or implementation of the complete 5G core or RAN components.

Work based on emulation can provide interesting results regarding the behavior of FL over wireless networks. However, emulated testbeds can lack heterogeneity, due to most or all system components being virtualized. In many emulated systems, the practical limitations of actual distributed system (e.g., heterogeneous compute power, heterogeneous communication latency) cannot be realized. Studies that implement FL on real computing hardware typically lack real implementations of wireless technologies~\cite{emperical_iot, Lee_Shin_2022, jiang_model_2023}. While these studies provide valuable insight into the limitations and performance benefits of FL over wireless, they do not completely bridge the gap between emulation and physical implementation. Additionally, despite the abundance of studies implementing real 5G testbeds~\cite{bonati_colosseum_2021, yu_cosmos_2019, powder} or real FL applications~\cite{jiang_model_2023, emperical_iot}, there is limited work on the intersection of FL and 5G.

Real-world deployments of FL over wireless networks remain limited but are of critical importance. Thus, researchers have begun building FL systems on real hardware to understand the challenges involved in practical implementation and the resulting performance limitations. For instance, in~\cite{jiang_model_2023}, the authors propose a model pruning framework to reduce the overall model size and training time, while maintaining fast convergence time. The authors implement and validate their framework on Raspberry Pi devices. In~\cite{Lee_Shin_2022}, the authors demonstrate FL over private 5G networks using the NWDAF. They configure a distributed NWDAF environment using Free5GC---an open-source 5G core implementation---showing how FL can be integrated within an open 5G testbed. 
In~\cite{emperical_iot}, the authors present a standard FedAvg system using the Flower framework to evaluate performance over small and large-scale IoT deployments with device heterogeneity. They assess test accuracy, convergence time, resource utilization, training time, and average model update exchange time between node and server.
This work aligns most with ours as they evaluate FL's communication limitations. A key difference is that the communication network in~\cite{emperical_iot} is implemented using rate-limited Ethernet connections, as opposed to wireless interfaces, thus limiting the applicability of the conclusions to FL over-the-air deployments.

\subsection{Main contributions}
In this work, we deploy a FL application using an open source 5G-NR SA testbed using OAIBox software, COTS SDRs and UEs, and the Flower AI FL Framework~\cite{flower_fl}. We developed comprehensive instrumentation tooling, which enabled the collection of diverse communication and machine learning performance metrics, including: FL model accuracy, computation time (e.g., server aggregation time, FL model training/evaluation times), transmission times (e.g., uplink/downlink times), and physical layer network metrics (e.g., SNR, BLER, CQI, etc.). We automated the experiment execution and collection process, which allowed us to compile a comprehensive dataset of FL-over-5G measurements. For reproducibility, we have open-sourced our dataset, FL application, instrumentation tools, and testbed configuration. 

Our main contributions are:
\begin{enumerate}
    \item Compile a comprehensive measurement dataset with 150 FL-over-5G experiments, which were automated using a custom containerized version of Flower. All data will be released. All source code was released on GitHub.
    \item Deploy a FL system using Flower AI, with custom latency metrics collection, which enables us to gather machine learning and communication metrics over time irrespective of the communications network being used.
    \item Evaluate common assumptions in FL-over-wireless literature using real-world measurements.
\end{enumerate}

This work utilizes the testbed and dataset described in Section~\ref{sec:implementation} to evaluate common assumptions in existing FL-over-wireless literature. Specifically, we evaluate throughput-proportional latency scaling, straggler identity, heterogeneity across rounds and devices, wall-clock timing assumptions, idealized throughput assumptions, and the persistence of these behaviors under congestion or on commercial networks.
\section{Background} \label{sec:background}
This work involves two adjacent specialties. This section provides background on the relevant components of both FL and 5G networks.

\subsection{Federated Averaging (FedAvg) Overview}
In FedAvg, each node $c \in \left\{1,...,N\right\}$ maintains a local dataset $\mathcal{D}_c$ and a machine learning model with weights and biases $w_c^t$. Upon initialization, the central server, called an aggregator, sets the initial model weights to random values, with $w_1^0 = \cdots = w_N^0$. These initial parameters are then transmitted to each participating node through the communication network. At each communication round $t$, a nonempty subset of nodes $\mathcal{C}$ with fraction $f$, where $|\mathcal{C}| = f\cdot N$, performs Stochastic Gradient Descent (SGD) on its local dataset for $E$ local epochs with batch size $B$ and learning rate $\eta$. The number of local SGD updates performed by each participating node with $|\mathcal{D}_c|$ samples is given by $u_c = \lfloor E|\mathcal{D}_c|/B \rfloor$, where the local dataset $\mathcal{D}_c$ is partitioned into batches of size $B$. After each local training round, the updated weights, $w_c^{t+1}$, are sent back to the server for aggregation using a weighted average. The server computes and then distributes the new global model $\phi_{t+1}$, defined as
\begin{equation*}
    \phi_{t+1} = \sum_{c\in\mathcal{C}}\frac{|\mathcal{D}_c|}{\sum_{j\in\mathcal{C}}|\mathcal{D}_j|}w_c^{t+1} \quad t \geq 0
    \label{eqn:fed_agg}
\end{equation*}
to all participating nodes, with $\phi_0 = w_N^0$. This cycle repeats until either the maximum rounds have been reached or the network converges to a loss/accuracy threshold.

\subsection{5G and O-RAN Overview}
A 5G network consists of the following components: end devices, Radio Access Network (RAN), which includes the gNodeB (gNB), and 5G Core Network (5GC). The end devices, known as User Equipment (UE), allow users to access the wireless network. The RAN consists of a radio unit and compute node that runs the base-station software and protocol stack. Finally, the 5GC coordinates among  the base stations, manages authentication, and establishes sessions between devices and external networks.

The 5G NR physical layer (PHY), as specified by 3rd Generation Partnership Project (3GPP)~\cite{3GPP-TS-38.104}, currently defines seven indexed numerologies defined by
\begin{equation}
    \Delta f = 2^\mu \cdot 15~\si{\kilo\hertz} \; ,
    \label{eqn:numerology}
\end{equation}
where $\Delta f$ is the Sub-Carrier Spacing (SCS). Regardless of the selected numerology, each radio frame is of duration $10~\si{\milli\second}$, which is divided into $10$ sub-frames of duration $1~\si{\milli\second}$ each. However, each sub-frame may contain a variable number of slots depending on the selected numerology. The numerology, specified by $\mu \in \{0, 1, \ldots, 6\}$, determines the SCS via \eqref{eqn:numerology}.

From this numerology configuration we can look at the Time-Division Duplex (TDD) split configuration\footnote{Only available while operating in TDD mode.}, and the channel bandwidth. A TDD configuration has the following parameters: periodicity, uplink slots, downlink slots, uplink symbols, and downlink symbols. The possible periodicity value that can be used for a given system is determined by the SCS. Using the combination of SCS and periodicity, the number of slots per period can be determined. The number of total slots per period can be any arrangement of uplink, downlink, and flexible slots; this can be configured by the user and the 3GPP standard does not specify any specific TDD configurations. 

The communication requirements of FL may align with the 5G capabilities. However, it is important to note that commercial mid-band NR deployments are often configured with DL-heavy TDD patterns, as is our OAI/OAIBox deployment under the tested configurations. Moreover, commercial radios follow that design and physical implementations using open-source components may include limitations when compared to commercial deployments, which we address in this paper.
\section{Implementation} \label{sec:implementation}
\begin{figure}
    \centering
    \includegraphics[width=\columnwidth]{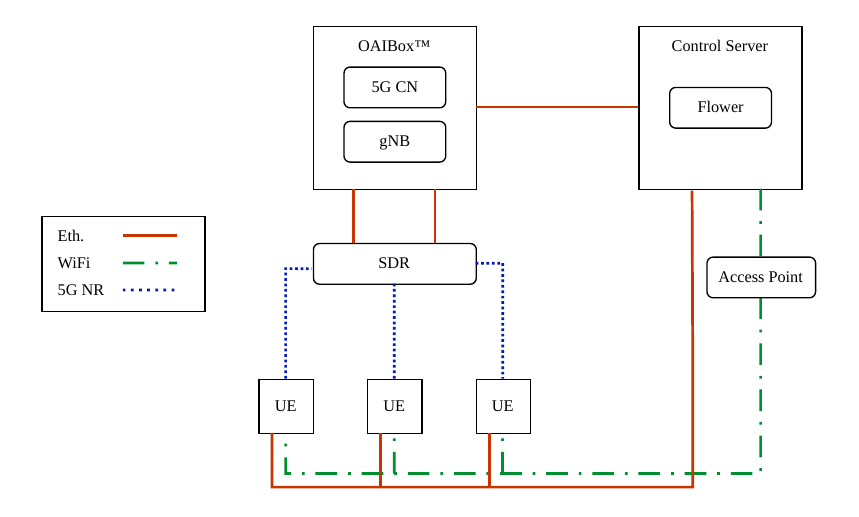}
    \caption{The communication network architecture showing the Ethernet, WiFi, and 5G links in communication between UEs and the control server.}
    \label{fig:network_arch}
\end{figure}

\begin{figure}[t]
    \centering
    \includegraphics[width=\columnwidth]{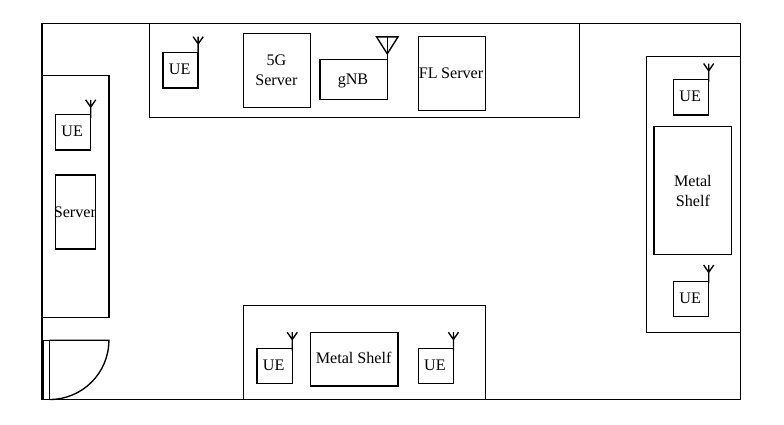}
    \caption{Blueprint of the Cellular Lab at our Northwestern University}
    \Description{A blueprint of the celluar lab at our Northwestern University. Showing a rectangular room with the folllowing components: 6 UEs, 1 5G Server, 1 FL Server, and a gNB.}
    \label{fig:blueprint}
\end{figure}

We consider a configurable communication network consisting of six UEs and a central control server, as shown in Fig.~\ref{fig:network_arch}. The network is configurable in that the nodes can connect to the control server via Ethernet (red), WiFi (green), or the 5G network (blue) via a Software Defined Radio (SDR) and an ORAN OAIBox Server. The UEs in the testbed are distributed throughout the room, ensuring spatial heterogeneity as shown in Fig.~\ref{fig:blueprint}. 

\subsection{Federated Learning Implementation}
\begin{figure}[t]
    \centering
    \includegraphics[width=1\linewidth]{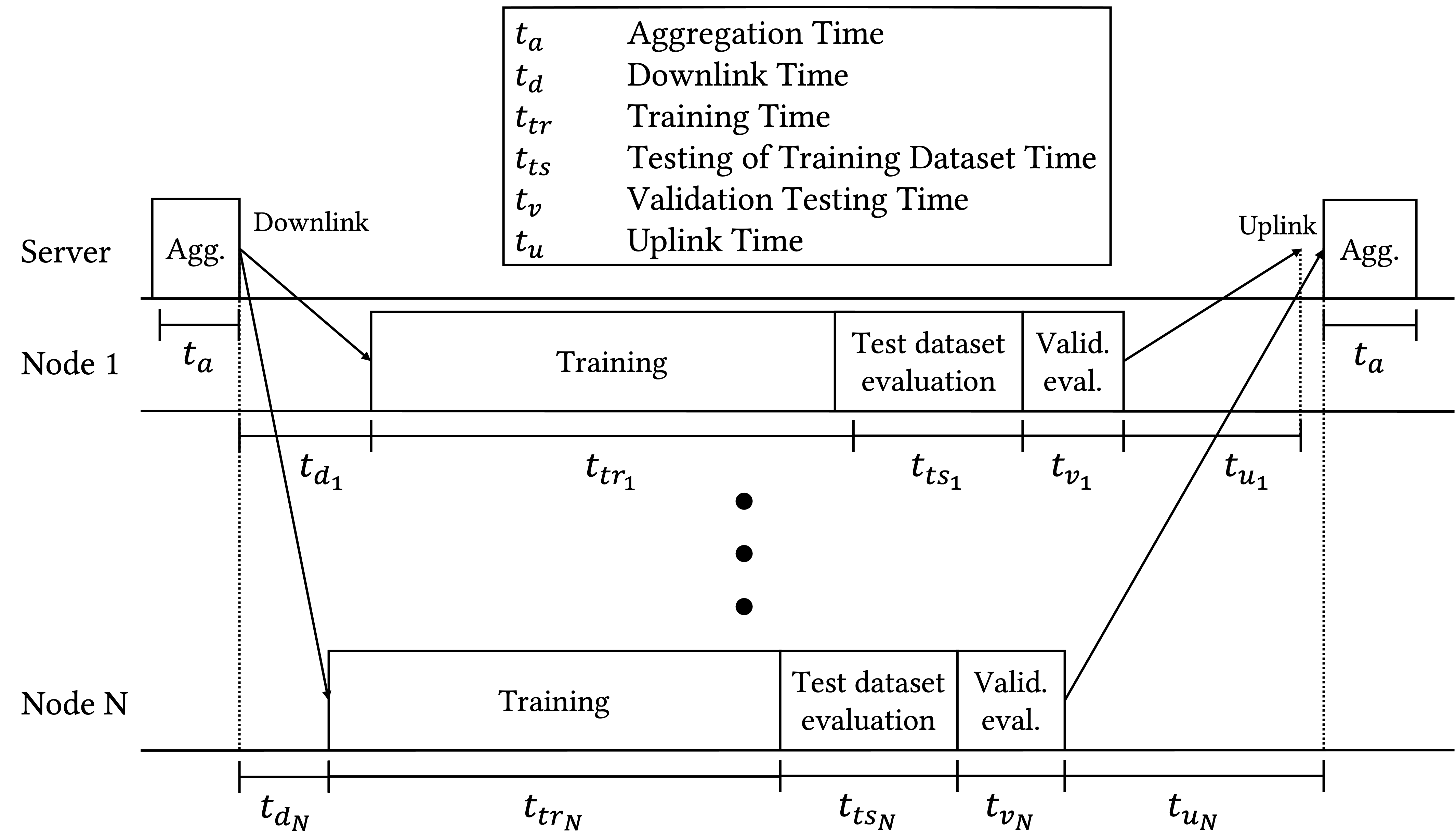}
    \caption{One Communication Round of Federated Learning}
    \label{fig:FL_diagram}
\end{figure}

Our system uses a distributed edge computing architecture with resource-constrained devices. A central server coordinates FL with $N$ nodes, communicating via an arbitrary network and is not dependent on the physical or data link layers. Fig.~\ref{fig:FL_diagram} illustrates one communication round of our FL application. The network includes six Raspberry Pis, without graphics acceleration processors, which serve as federated learning nodes. Each node has its own partition of a global image dataset, which is partitioned and distributed by the server. This application utilizes Flower FL---a communication agnostic framework---that provides \emph{``A unified approach to federated learning, analytics, and evaluation''~\cite{flower_fl}}. Flower enables users to federate an arbitrary machine learning model, dataset, and FL strategy, while providing extensibility for customized evaluation and metrics gathering. We use the FedAvg algorithm~\cite{mcmahan_communication-efficient_2017} for parameter aggregation. For ease of orchestration, all FL applications run in a containerized environment. All containers communicate with each other using manually created virtual networks.

The primary model in the network is SqueezeNet~\cite{SqueezeNet}, due to its low training time and relatively low model weight size ($2.9172$~\si{\mega\byte}). We use the hyperparameters recommended by the TorchVision implementation~\cite{torchvision_classification_scripts} and listed in Table~\ref{tab:fl_params}. 

We perform image classification using the CIFAR-10 dataset~\cite{cifar10} which is distributed using the Dirichlet dataset partition method~\cite{dirichlet} with a concentration parameter $\alpha=1$. Image classification is a well-established task for federated learning~\cite{lim_federated_2020, banabilah_federated_2022}. Therefore, our measurement methodology is generalizable to any FL task as the communication method is agnostic of FL complexity. More complex models only influence the ratio of compute to communication time.

\begin{table}[t]
    \caption{FL Hyper-parameters}
    \centering
    \begin{tabular}{lc}
        \toprule
        \textbf{Hyperparameter} & \textbf{Value} \\
        \midrule
        Dataset & CIFAR-10~\cite{cifar10} \\
        
        Model & SqueezeNet~\cite{SqueezeNet} \\
        
        Local Epochs  & 1 \\
        
        Batch Size & 128 \\
        
        Momentum & 0.9 \\
        
        Learning Rate & 0.01 \\
        
        Weight Decay & 0.0002 \\
        \bottomrule
    \end{tabular}
    \label{tab:fl_params}
\end{table}

\subsection{Network Implementation}\label{sec:network_arch}
As previously stated, each device in our network has access to Ethernet, WiFi, and the ORAN 5G testbed. Each node on the WiFi is connected to the same access point, with the configuration parameters listed in Table~\ref{tab:wifi} acquired from the network settings reported by Linux.

The 5G ORAN infrastructure includes two servers: one to run the OAI core network, and the other to run the OAI monolithic gNB. These servers interface with each other through the institution's Ethernet, which has a throughput well above the capability of the 5G network. Each Raspberry Pi interfaces via USB 3.0 to a Telit 980m 5G modem to provide access to the 5G network. Note that the use of these USB cellular modems may incur some additional latency, compared to the standard 5G chipset integrated with the device.

Our network's link parameters were measured using the iperf utility, and the results are listed in Table~\ref{tab:links}. The table shows the link speed and TCP burst latency\footnote{TCP burst latency is measured by iperf and includes socket write times, buffer delay, network transport time, and socket read times. This is not the same as round-trip time (RTT).\label{fn:latency}} for the connection between the central server to the nodes on Ethernet, WiFi, and 5G, and the link parameters of the Ethernet connection between the gNB and 5GC.

The RAN operates in band n78 in stand-alone mode with a 30~\si{\kilo\hertz} SCS. Additionally, we have the ability to easily configure the RAN with the following TDD ratios (DL:UL): 7:2, 5:4, 2:7, 3:1, 2:2. Further, we are able to select from the following bandwidths: 100, 80, 60, 40, 20 MHz. The gNB uses OAI software connected to a USRP X310 SDR via 10Gb Ethernet interfaces, with an internal clock reference for timing synchronization. The antennas of our USRP are omnidirectional. This is compliant with the asymmetric design of 5G-NR. 

Regardless of bandwidth, the system operates with Synchronization Signal Block (SSB) Absolute Radio-Frequency Channel Number (ARFCN) at 620544 and Point A of 620016 with a center ARFCN of 622620. Transmit path is configured with 26~dB attenuation, maximum PDSCH reference signal power is set to -27~dBm, and the max receive gain is 32~dB. PRACH uses a configuration index 148 to accommodate the 30~kHz subcarrier spacing. 
The physical layer parameters of the gNB are shown in Table~\ref{tab:5g_testbed}, and they are optimized for stability of the testbed.
\begin{table}[t]
    \caption{WiFi node Physical Layer Parameters}
    \centering
    \begin{tabular}{lc}
        \toprule
        \textbf{Parameter} & \textbf{Value} \\
        \midrule
        Frequency  & $5.62~\si{\giga\hertz}$\\
        Channel & $124$ \\
        Bandwidth & $20~\si{\mega\hertz}$ \\
        Signal Strength & $-51~\;\si{\decibel\meter}\pm5~\si{\decibel\meter}$ \\
        \bottomrule
    \end{tabular}
    \label{tab:wifi}
\end{table}

\begin{table}[t]
    \centering
    \caption{Testbed Capacity by Connection Path and Medium (measured from device 1 over a 30-second iperf trial)}
    \resizebox{\columnwidth}{!}{%
    \begin{tabular}{llcccc}
        \toprule
        \multirow{2}{*}{\textbf{Path}} & 
        \multirow{2}{*}{\textbf{Medium}} & 
        \multicolumn{2}{c}{\textbf{TCP Burst Latency\footref{fn:latency} (\si{\milli\second})}} & 
        \multicolumn{2}{c}{\textbf{Throughput (\si{\mega\bit\per\second})}} \\
        \cmidrule(lr){3-4} \cmidrule(lr){5-6}
        & & \textbf{Uplink} & \textbf{Downlink} & \textbf{Uplink} & \textbf{Downlink} \\
        \midrule
        \multirow{3}{*}{Server$\leftrightarrow$Node} & 
        WiFi & 290.4 & 30.444 & 95.0 & 72.2 \\
        & Ethernet & 18.298 & 3.723 & 936 & 934  \\
        \midrule
        \multirow{3}{*}{Core$\leftrightarrow$Node} & 
        5G (40 MHz, 7:2 TDD) & 1100.9 & 284.4 & 22.9 & 79.8 \\
        &5G (40 MHz, 2:7 TDD) & 338.7 & 384.7 & 87.5 & 63.7 \\
        &5G (100 MHz, 7:2 TDD) & 845.9 & 296.4 & 28.3 & 85.1 \\
        &5G (100 MHz, 2:7 TDD) & 303.0 & 332.9 & 107 & 71.5 \\
        \midrule
        Server$\leftrightarrow$gNB & Ethernet & 0.485 & 0.174 & 2350 & 2350 \\
        \bottomrule
    \end{tabular}%
    }
    \label{tab:links}
\end{table}

\begin{table}[t]
    \centering
    \caption{5G Testbed Configuration Details}
    \label{tab:5g_testbed}
    \resizebox{\columnwidth}{!}{%
    \begin{tabular}{ll}
        \toprule
        \textbf{Parameter} & \textbf{Value} \\
        \midrule
        3GPP Specification & Rel 16 \\

        Operating Mode & SA \\
        
        Operating Band & n78 \\
        
        Software & OAIBox OpenAirInterface \\
        
        gNodeB & USRP X310 SDR \\

        User Equipment (UE) & Telit 980m \\
        
        Interface & 2x10G SFP+ \\
        
        Clock Reference & Internal \\

        Antenna & Omnidirectional \\
        \midrule
        Flexible Slot & 6 DL, 4 Guard, 4 UL [symbols]\\
        
        SSB ARFCN & 620544 (3308.160 MHz) \\
        
        Point A ARFCN & 620016 (3300.240 MHz)\\

        Center ARFCN & 622620 (3339.3 MHz) \\
        
        TX, RX Attenuation &$ 0~\si{\decibel}, 26~\si{\decibel}$\\

        Max RX Gain & 32~\si{\decibel} \\
        
        Max PDSCH Reference Signal Power & $-27~\si{\decibel\meter}$ \\
        
        PRACH Configuration Index & 148 \\
    
        \bottomrule
    \end{tabular}%
    }
\end{table}

\subsection{Instrumentation} \label{sec:monitory}
We implement instrumentation tooling which allows us to capture the computation time, communication time, and physical layer values in a FL server round. We define computation time as the time to train, evaluate, and aggregate the model.
We collect communication time and all computation time measurements through the Flower framework. Communication time is defined as the sum of both uplink and downlink model transmission latency. To measure the model transmission time we modify the Flower framework source code to inject local timestamps on each node. Within Flower, there are two methods the application uses for transmission: \texttt{send} and \texttt{receive}. Both utilize the gRPC~\cite{grpc2024} application layer protocol to transfer parameters---these functions wait for completion before exiting. Using this, we can extract the model weight uplink and downlink transmission times. We observe a $56.56~\si{\milli\second}$ overhead which is negligible given the second-scale measurements.

Lastly, we extract physical layer data from the gNB, which is outputted as a database of Radio Network Temporary Identifiers (RNTI) to their respective physical layer measurements. Due to Radio Resource Control (RRC), re-establishment during the experiment and the temporary nature of RNTIs, the database includes numerous RNTI values per UE. Thus, correlating the physical layer measurements to the FL values is not trivial. In some experiments, including the ones for which we discuss physical layer metrics in this paper, we determined the identity of each device by tracking the order in which each began uplink transmission from the FL metrics and correlating with uplink throughput peaks from the PHY data across the 200 rounds of each experiment.

Data analysis is performed using custom Python scripts. In summary, we compile a database containing uplink time, downlink time, train time, evaluation time, train loss, evaluation loss, evaluation accuracy, wall-clock time, and physical layer measurements (PCMAX, PUSCH SNR, PHR, uplink/downlink MCS and modulation index, SINR, PUCCH SNR, RSRQ, RSSI, CQI, uplink/downlink BLER and carrier frequency).

\section{Experimental Results and Discussion}\label{sec:results}
\begin{figure}[t]
    \centering
    \includegraphics[width=\columnwidth]{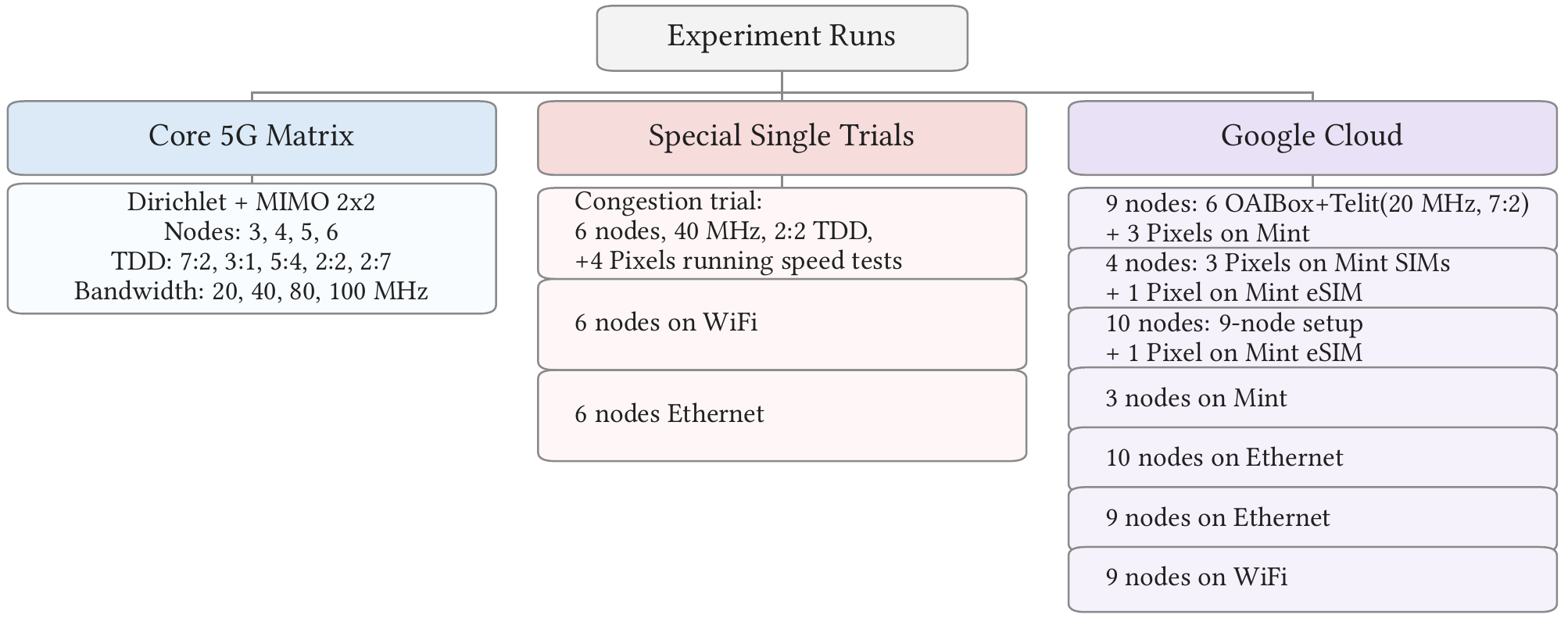}
    \caption{Overview of experiment scenarios for Ethernet, WiFi, and 5G under different configurations.}
    \label{fig:flowgraph}
\end{figure}

We present a comprehensive FL-over-5G dataset, comprising over 350 hours of measurements from 150 FL experiments. We fixed the random seeds for dataset distribution and FL training to ensure reproducibility and to isolate the effects over 5G configurations, which is a main object of our study. All experiments ran for 200 FL rounds with no early stopping. Figure~\ref{fig:flowgraph} outlines the collected measurements for FL over Ethernet, WiFi, and 5G with varying configurations (namely bandwidth, TDD, and number of nodes) using 2x2 MIMO and Dirichlet distribution. In addition, we conducted a congestion experiment with ten nodes: six performed FL while the remaining four generated background traffic. We also measured the FL performance when the FL server was deployed in the cloud over a mixed network of the OAI testbed and Mint Mobile as a commercial 5G deployment. Finally, we collected data for the FL task using six nodes connected over Ethernet and WiFi. In line with open science principles, all collected data, including final trained models and 5G physical layer measurements compiled will be made available for download. In this section, we provide an overview of the most significant results. 

\subsection{Network Comparison}
According to our measured throughput in Table~\ref{tab:links}, the Ethernet link is approximately ten times faster than WiFi. The WiFi uplink is approximately four times faster than 40 MHz, 7:2 TDD 5G, with roughly equal downlink speeds. In Fig.~\ref{fig:comm_round_over_network}, we show communication and overall round time over each of these networks. We observe a mean downlink time of $0.22\si{\second}$ over Ethernet, $0.89\si{\second}$ over WiFi, and $1.19\si{\second}$ over 5G and uplink time of $0.23\si{\second}$, $0.63\si{\second}$, and $4.40\si{\second}$, respectively. Rather than a ten-fold improvement from WiFi to Ethernet, we observe a factor of 2.7 (uplink) and 4 (downlink). Comparing 5G to WiFi, the uplink time diverges from what the single-node throughput predicts. Comparing Table~\ref{tab:links} to Fig.~\ref{fig:comm_round_over_network}, we observe that downlink time is four times higher than the throughput predicts for WiFi and seven times higher for 5G. Main factors contributing to these contrasts are WiFi contention, 5G multi-user scheduling, and network control overhead. Effects of 5G multi-user scheduling will be discussed in Fig.~\ref{fig:5N_6N_40MHz_2-2_throughput}.

Notably, Fig.~\ref{fig:comm_round_over_network} shows that \textbf{communication time accounts for approximately 26.8\% of the overall round time in the 5G configurations compared to 9.0\% for WiFi and 2.9\% for Ethernet}, highlighting the importance of communication time for FL tasks (especially over wireless networks).

\subsection{Wall-Clock Time versus Number of Rounds}

\begin{figure}[t]
    \centering
    \includegraphics[width=1\linewidth]{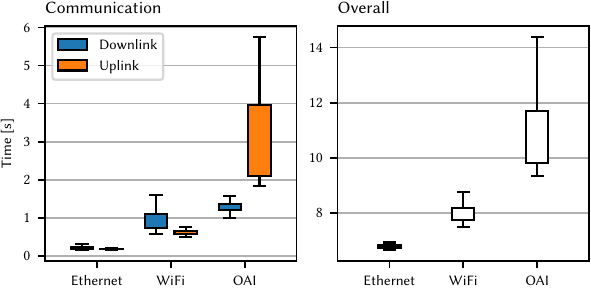}
    \caption{Communication time and overall round time over Ethernet, WiFi, and 5G (40 MHz, 7:2 TDD) with 6 nodes.}
    \label{fig:comm_round_over_network}
\end{figure}

\begin{figure}[t]
    \centering
    \includegraphics[width=1\linewidth]{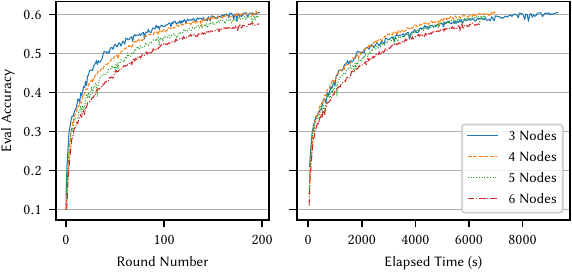}
    \caption{Evaluation accuracy for an increasing number of nodes over round number (left) and over wall-clock time (right) on 40 MHz 5G with 2:2 TDD.}
    \label{fig:accuracy_rounds_time}
\end{figure}

Existing literature that seeks to lower convergence time of the FL task, e.g.,~\cite{mcmahan_communication-efficient_2017, jiang_model_2023, sad-async-peertopeer, mu-split}, often focuses on number of rounds or the number of model parameter exchanges. This neglects the effects of FL algorithm, number of nodes, and model size on computation and communication time, which as shown in Fig.~\ref{fig:comm_round_over_network}, can have a significant impact on wall-clock convergence time.

Figure~\ref{fig:accuracy_rounds_time} shows the average validation accuracy over rounds (left) and over wall-clock time (right) as we increase the number of nodes ($N$). From Sec.~\ref{sec:background} and, originally, in~\cite{mcmahan_communication-efficient_2017}, it follows that increasing the number of nodes, $N$, will affect validation accuracy \emph{over rounds}. Specifically, increasing $N$ decreases the size of the local dataset, which decreases the number of FL updates per round $u_c$, resulting in a reduction on the accuracy at a fixed number of rounds, which largely agrees with Fig.~\ref{fig:accuracy_rounds_time} (left). In contrast, the \textbf{evolution of validation accuracy over \emph{wall-clock time} in Fig.~\ref{fig:accuracy_rounds_time} (right), which accounts for the impact of number of nodes $N$ on computation time and communication time, can potentially lead to different results}. This is because increasing the number of nodes $N$ increases the communication time (due to 5G multi-user scheduling) and decreases the computation time (due to the lower number of FL updates per round) leading to the more complex evolution in Fig.~\ref{fig:accuracy_rounds_time} (right).

\subsection{Heterogeneity}

\begin{figure*}[h]
    \centering
    \includegraphics[width=1\linewidth]{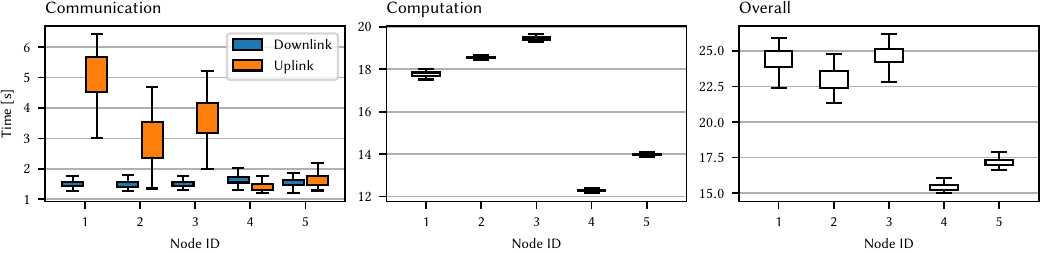}
    \caption{Per-device communication, computation, and overall round time for 5G with 5 nodes, 40 MHz bandwidth, 2:2 TDD.}
    \label{fig:5N_40 MHz_2-2}
\end{figure*}

\begin{figure}[t]
    \centering
    \includegraphics[width=\columnwidth]{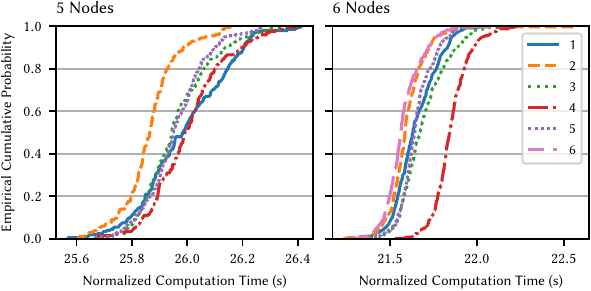}
    \caption{Empirical CDF of computation time, normalized to number of samples per device, for five nodes (left) and six nodes (right) over 5G with 40 MHz bandwidth and 2:2 TDD.}
    \label{fig:atyp_typ_norm_comp}
\end{figure}
Common communication/computation homogeneity assumptions in the literature include that communication time is equal across devices~\cite{Alanazi-fedband,Zehtabi-decentralized-sporadic, avdyukhin_federated}, communication time is stationary~\cite{fraboni_general,narmadha_fedeff_2025,chen-fsreal,skocaj_uplink_2023}, and computation time is parameterized by hardware attributes (CPU clock frequency, mini-batch size, etc.)~\cite{ma_adaptive_2023,wan_convergence_2021,narmadha_fedeff_2025}. For example, in~\cite{Zehtabi-decentralized-sporadic, avdyukhin_federated}, the authors assume that communication is randomly available, but the communication time is constant among participating clients when available; and in~\cite{narmadha_fedeff_2025,chen-fsreal}, the authors assume that clients' transmission rates are uniformly distributed, but remain constant over the experiment.

In Fig.~\ref{fig:5N_40 MHz_2-2} we show the communication, computation and overall round time measured by each device for the 40 MHz, 2:2 TDD 5G configuration with five nodes with identical hardware. We observe that communication time not only varies by device but also over time, with uplink varying by several seconds for some devices. \textbf{Main reasons for the uplink communication time variability are wireless channel dynamics and 5G multi-user scheduling}. It can be seen from Fig.~\ref{fig:5N_40 MHz_2-2} that node with ID 4 is often the first to finish training and start uplink model transmission by itself. In contrast, when node with ID 3 finishes training, it will often transmit its model concurrently with nodes with IDs 1 and 2, which contributes to their increased uplink communication time. 

In Fig.~\ref{fig:5N_40 MHz_2-2}, we also observe that computation time varies between devices. This is mainly due to the non-uniform (Dirichlet) dataset distribution across devices, wherein each node trains on a different amount of examples, skewing the training time. For a fairer comparison of computation time, in Fig.~\ref{fig:atyp_typ_norm_comp} we show the compute time \emph{normalized to the number of local examples}. \textbf{We observe that after normalization, the computation time becomes more similar, as expected, but it is still not identical across devices nor stationary in time, despite the identical hardware}. 

\subsection{Identity of the Straggler}
\begin{figure}[t]
    \centering
    \includegraphics[width=1\columnwidth]{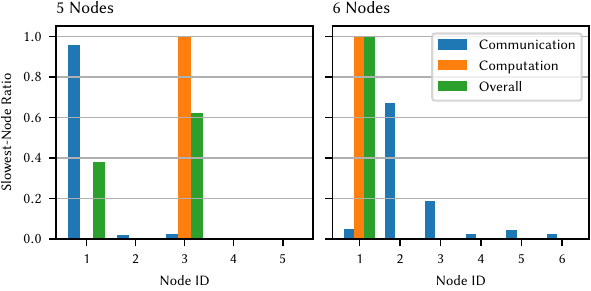}
    \caption{Fraction of rounds each device was the slowest in communication, computation, and overall round completion (i.e., the straggler) for five nodes (left) and six nodes (right) over 5G with 40 MHz bandwidth and 2:2 TDD.}
    \label{fig:atyp_typ_straggler}
\end{figure}

\begin{figure}[t]
    \centering
    \includegraphics[width=\linewidth]{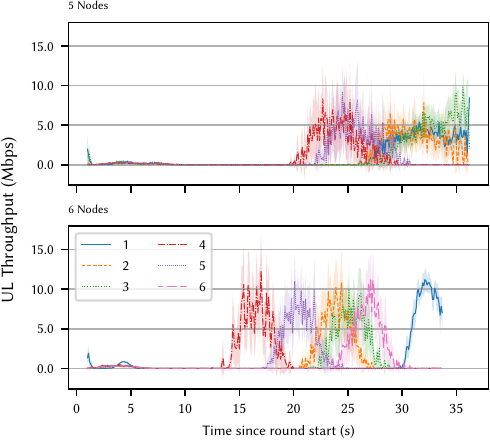}
    \caption{Per-node uplink throughput over a single round, calculated from PHY-layer metrics for five (top) and six (bottom) nodes over 5G with 40 MHz bandwidth and 2:2 TDD.}
    \label{fig:5N_6N_40MHz_2-2_throughput}
\end{figure}

\begin{figure}[t]
    \centering
    \includegraphics[width=1\columnwidth]{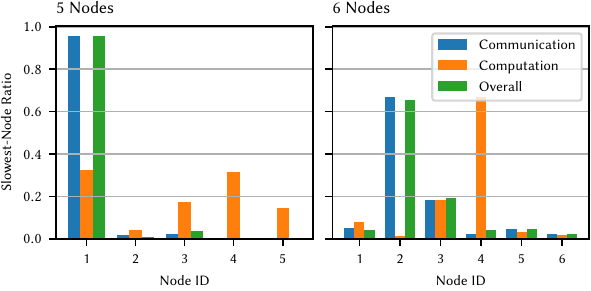}
    \caption{Fraction of rounds each device was the slowest in communication, computation, and overall round, for five nodes (left) and six nodes (right) over 5G with 40 MHz bandwidth and 2:2 TDD, with computation time normalized to number of samples per device.}
    \label{fig:atyp_typ_norm_comp_straggler}
\end{figure}

While the assumption of stationary communication and computation time portends a consistent straggler, we show that this does not hold in our experiments. In Fig.~\ref{fig:atyp_typ_straggler}, we show the fraction of time each device was the slowest in communication, computation, and overall round completion (typically called the straggler) for five and six nodes over 5G with 40 MHz bandwidth and 2:2 TDD. With six nodes (right), we see that there is a consistent straggler. We consider an experiment to have a consistent straggler when the same node was the last to complete its round in at least 95\% of the 200 rounds. \textbf{In all of our experiments with consistent stragglers, the slowest-computation node was the overall straggler}. 

Notably, \textbf{only 70\% of our experiments had consistent stragglers. In the other 30\%, the straggler was neither consistent nor uniformly-distributed, with distributions similar to Fig.~\ref{fig:atyp_typ_straggler} (left)}. Competing stragglers often have similar computation times, which leads to concurrent uplink communications, and the straggler competition being determined by the node that finishes uplink first. This competition can be observed in Fig.~\ref{fig:5N_40 MHz_2-2} (right).

In Fig.~\ref{fig:5N_6N_40MHz_2-2_throughput}, we display average per-node throughput over the duration of a single round. Throughput is calculated from PHY-layer metrics tracking the number of bytes each node sends per second. Throughput is averaged over all rounds. In Fig.~\ref{fig:5N_6N_40MHz_2-2_throughput}, we can observe concurrent uplink transmissions. With six nodes (bottom), node with ID 1 lagged behind in computation far enough that by the time it began uplink, it had a clear channel. Conversely, with five nodes (top), nodes with IDs 1, 2, and 3 all transmit at the same time with reduced throughput. While node 3 is the slowest to train, it is not offset much from nodes with IDs 1 and 2, resulting in their concurrent uplink transmissions and leading to an inconsistent straggler. The gap between the slowest node and the previous ones, rather than just the ordering of completing training, helps determine the straggler.

Because computation time is skewed by the non-uniform (Dirichlet) dataset distribution across devices, we expect that the probability of concurrent uplink transmissions is reduced, and therefore the probability of competing stragglers is reduced. To unskew computation time, we show in Fig.~\ref{fig:atyp_typ_norm_comp_straggler} the slowest-node ratio with computation time normalized to the number of samples per device, using the same 5G configuration as in Fig.~\ref{fig:atyp_typ_norm_comp}. Interestingly, while Fig.~\ref{fig:atyp_typ_straggler} (with skewed computation time) showed a consistent straggler for the six node case and competing stragglers for the five node case, our unskewed results in Fig.~\ref{fig:atyp_typ_norm_comp} suggest the opposite. In both cases, we observe a more even, though not uniform, distribution of the straggler node.

\subsection{Comparison to Theoretical Rates}
\begin{figure}[t]
    \centering
    \begin{subfigure}[t]{\linewidth}
        \centering
        \includegraphics[width=\linewidth]{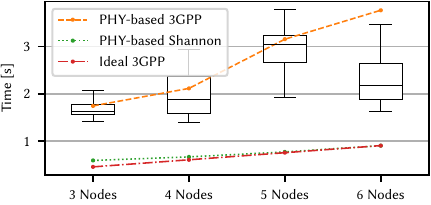}
        \caption{Varied number of nodes connected to 5G with 40 MHz bandwidth and 2:2 TDD configuration}
        \label{fig:40MHz_2-2_theoretical}
    \end{subfigure}
    \begin{subfigure}[t]{\linewidth}
        \centering
        \includegraphics[width=\linewidth]{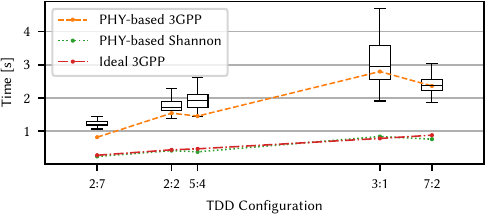}
        \caption{Six nodes connected via 80 MHz 5G with varying TDD configurations.}
        \label{fig:80MHz_6N_theoretical}
    \end{subfigure}
    \caption{Measured uplink time (box plots) compared to data-driven 3GPP model (orange, dashed), data-driven Shannon capacity (green, dotted), and ideal 3GPP with modulation order $Q_m=6$ (red, dash-dot).}
    \label{fig:theoretical}
\end{figure}

We calculate the expected uplink time for each experiment according to the 3GPP TS 38.306 specification and the Shannon capacity theorem. We calculate the instantaneous throughput indicated by each model using the relevant PHY data: uplink channel SINR for Shannon and modulation order ($Q_m$) and target coding rate ($R_x$) (indicated by uplink MCS index for 3GPP) at each data point. For each node, we calculated the percentage of the round spent in uplink transmission. Then, we averaged the model-based throughput over all nodes in each round. Finally, we divided the model data size by the model-based throughput and by the number of nodes over each round to find the expected uplink time, and we plot the median.
For the 3GPP model, we calculate both the PHY data-based uplink time and the time given by the ideal channel conditions ($Q_m=6$ with the maximum LDPC target coding rate). For the Shannon capacity, we only plot the PHY data-based value. In Fig.~\ref{fig:40MHz_2-2_theoretical}, we show the measured uplink time compared to the three models for a varying number of nodes connected via 40 MHz, 2:2 TDD 5G configuration. We observe that the data-driven 3GPP uplink times match more closely than the other two. Neither the ideal 3GPP nor the Shannon capacity accurately models the uplink time. In Fig.~\ref{fig:80MHz_6N_theoretical}, we show the measured and model-based uplink times for 6 nodes connected via 80 MHz 5G over varying TDD configurations. We observe the same effect: while the data-driven 3GPP modeled uplink time approaches the measured values, the ideal 3GPP and data-driven Shannon capacity do not come close.

Existing literature that accounts for the wireless channel, e.g.,~\cite{chen_convergence_2021,skocaj_uplink_2023,tran_federated_2019,wan_convergence_2021,pase-imperfect-csi}, often assumes that the model upload occurs at rates comparable with the Shannon capacity~\cite{shannon_capacity}. While these analyses are useful in understanding the relative channel effects and evaluating the performance in increasingly efficient networks, they may be inaccurate when compared with real-world networks. In our measured versus modeled uplink time plots, we show the importance of using real measurements or realistic, data-driven modeling in FL research.

\subsection{Real-World Considerations}
\subsubsection{Commercial Comparison}
\begin{figure}[t]
    \centering
    \begin{subfigure}[t]{0.49\columnwidth}
        \centering
        \includegraphics[width=\columnwidth]{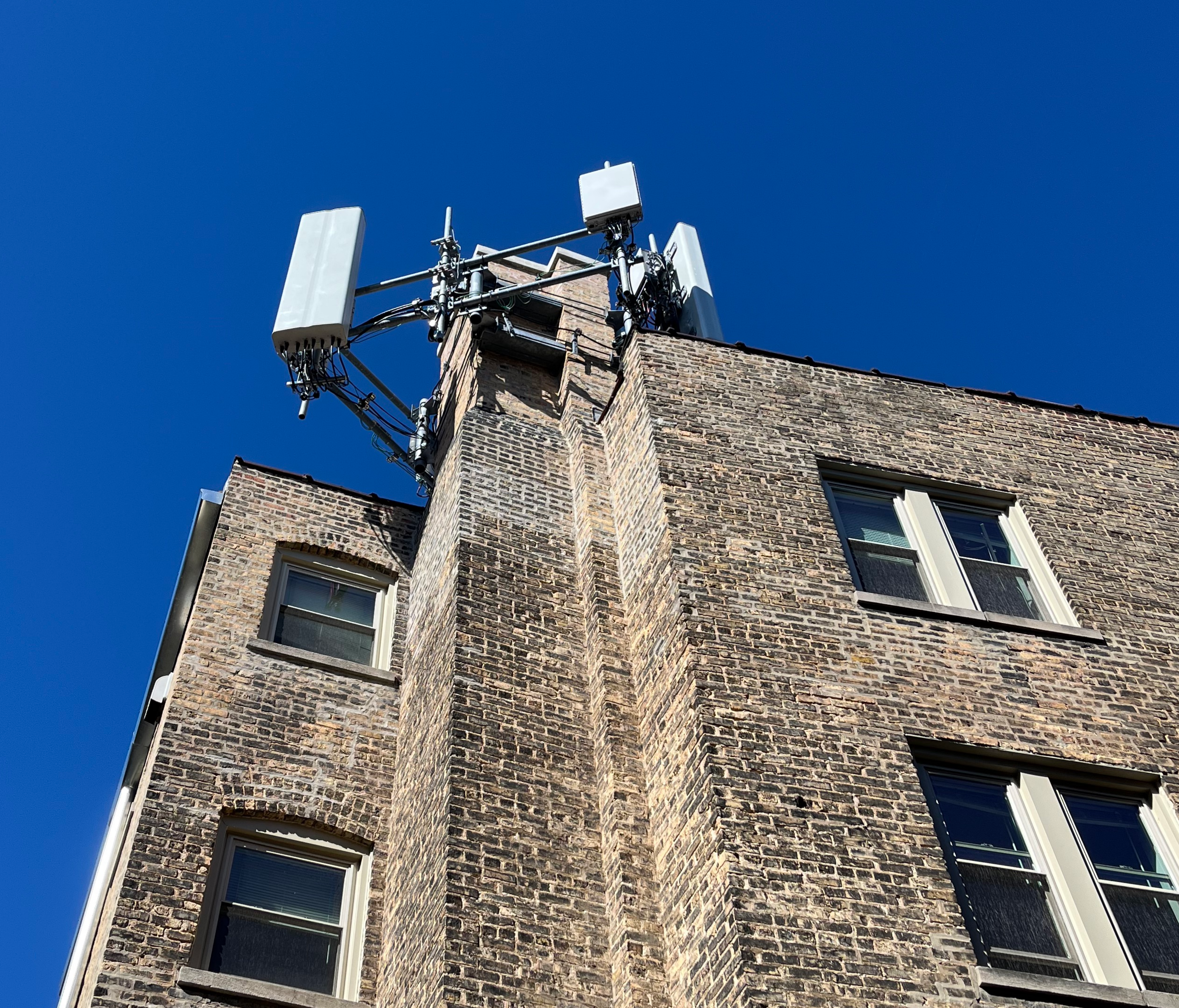}
        \label{fig:cell-tower}
    \end{subfigure}
    \begin{subfigure}[t]{0.45\columnwidth}
        \centering
        \includegraphics[width=\columnwidth]{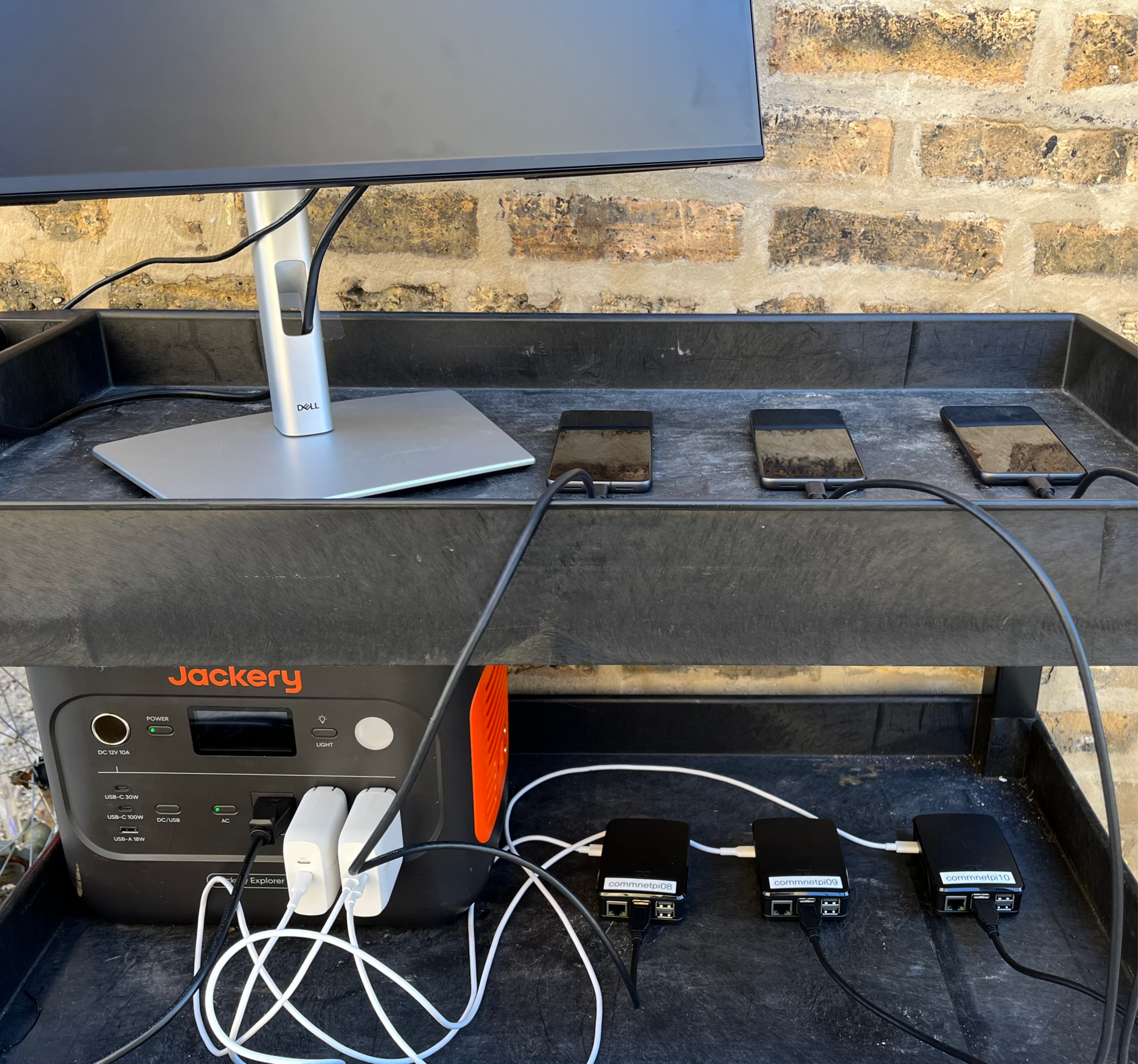}
        \label{fig:mint-outdoor}
    \end{subfigure}
    \vspace{-0.3cm}
    \begin{subfigure}[t]{0.85\columnwidth}
        \centering
        \includegraphics[width=\columnwidth]{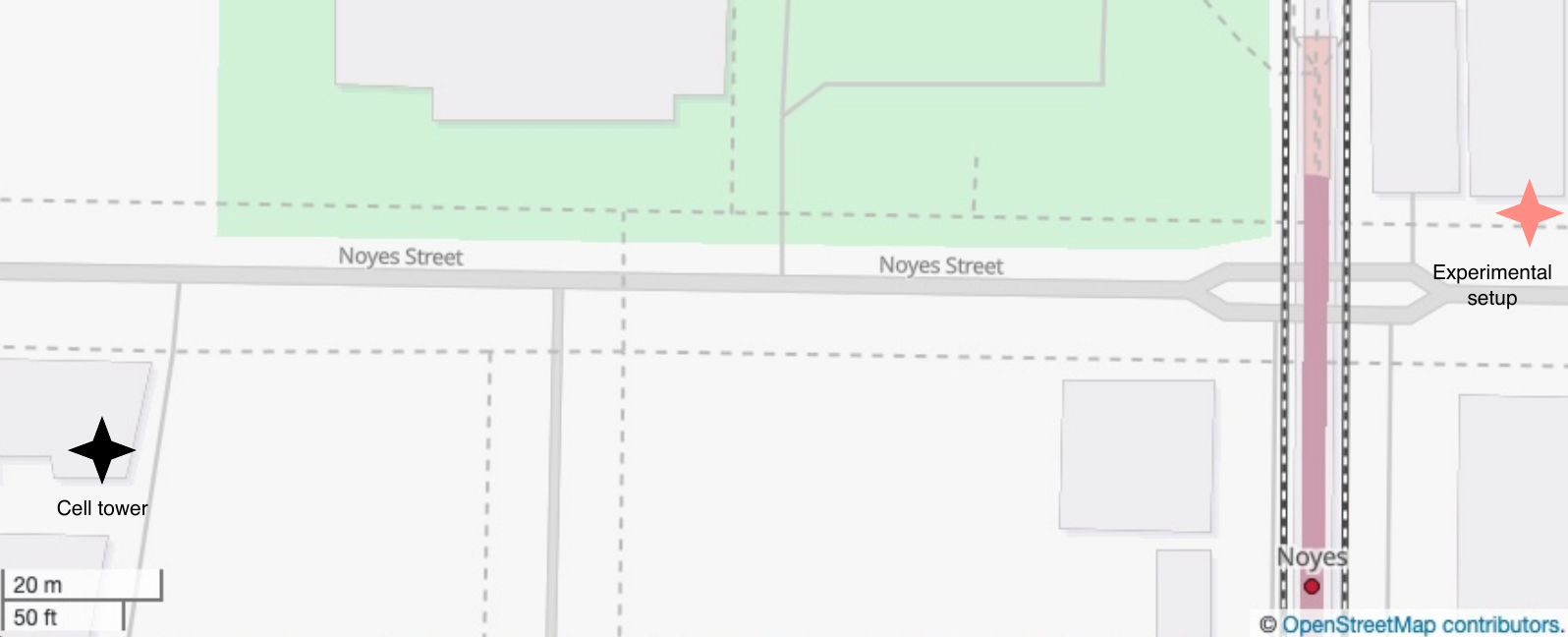}
        \label{fig:map}
    \end{subfigure}
    \vspace{-0.3cm}
    \caption{Outdoor experimental setup. Cell tower (top-left), mobile work station (top-right), and map of relative locations.}
    \label{fig:outdoor-experiments}
\end{figure}

\begin{figure*}[h]
    \centering
    \includegraphics[width=\linewidth]{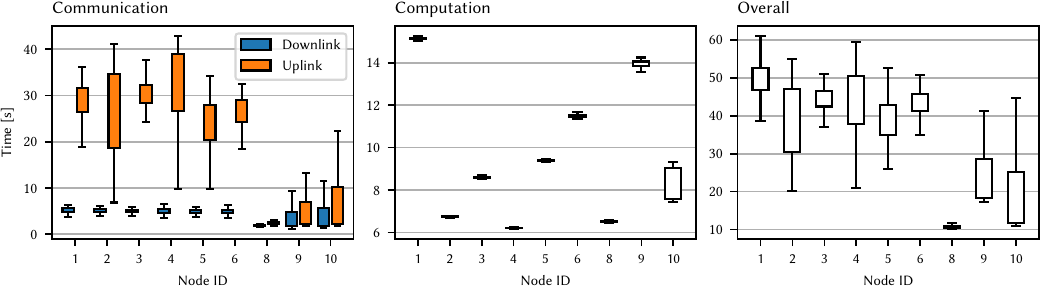}
    \caption{Communication, computation, and overall round time per node with 6 devices connected via OAIBOX (20 MHz, 7-2 TDD) and three devices connected via Mint Mobile.}
    \label{fig:mint-outdoor_comm_comp_round}
\end{figure*}

To relate our measurements to commercial 5G, we deployed our FL system over Google Cloud with three Raspberry Pis connected to Mint Mobile via USB-tethering on Google Pixel 8s. The majority of these experiments was performed indoors, and one experiment was outdoors with line of sight visibility to the cell tower. The outdoor (mixed-network) experiment was conducted with six Raspberry Pis connected to the OAIBox system and three connected to Mint Mobile (See Fig.~\ref{fig:outdoor-experiments}). We show the results of our mixed-network experiment in Fig.~\ref{fig:mint-outdoor_comm_comp_round}. We configured our network to best resemble a commercial network's needs and constraints, with high downlink capability over relatively low bandwidth (7:2 TDD, 20 MHz). The nodes connected over OAI performed worse than the nodes connected over Mint, but we note that our results are within an order of magnitude of the commercial network. Further, we chose the worst configuration of OAI to compare to Mint, and we had 6 devices connected, whereas we only connected three to Mint. We note that comparison to Mint is not a complete validation of our work, but demonstrates the open-source testbed performs similarly to commercial systems.

\subsubsection{Congestion}
\begin{figure}[t]
    \centering
    \includegraphics[width=1\linewidth]{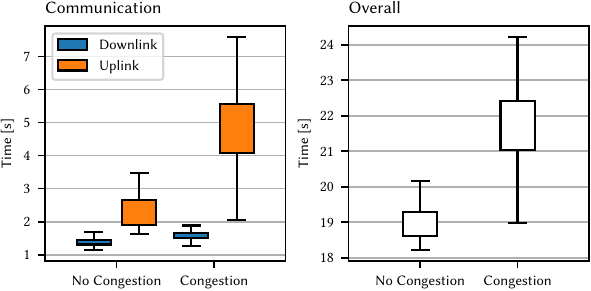}
    \caption{Communication and round time with and without external congestion in a six-node 5G network with 40 MHz bandwidth and 2:2 TDD configuration.}
    \label{fig:congestion}
\end{figure}
We investigated the effect of external network congestion on FL by connecting four Pixels to the 5G network. Two Pixels performed uplink speed tests over the trial's duration, while the other two performed downlink speed tests. In Fig.~\ref{fig:congestion}, we show communication and round time for the same 5G configuration with and without congestion. We observe that congestion nearly doubles communication time, increasing it from an average of $3.67\si{\second}$ to $6.56\si{\second}$.

\section{Conclusion and Future Work} \label{sec:conclusion}
We develop a measurement-focused 5G testbed and present a comprehensive FL-over-5G dataset to support future research and inform future system design. Our dataset includes measurements of FL over 5G in a wide range of configurations, varying bandwidth, TDD, MIMO, dataset distribution, and congestion. Additionally, we collected measurements of the performance of the FL application when using a commercial 5G network. Using this dataset, we evaluate common assumptions of FL over wireless. Specifically, we find that there is a consistent straggler in 70\% of trials, while in the remaining trials, high communication latencies cause competing stragglers. Future work will explore system performance with alternate FL strategies, scheduling optimizations, and increased model complexity. Additional research directions include implementing the aggregator in the NWDAF as simulated in literature, extending the testbed to Frequency Range 2/3 (FR2/FR3) frequencies to leverage their inherent performance benefits, and comparing purpose-built 5G networks, i.e., commercial deployments, to open-source solutions. In line with open science principles, all software developed for this paper is available on our \href{https://github.com/Net-X-Research-Group/federated_learning_testbed}{GitHub repository}. All collected data, including final trained models and 5G physical layer measurements, will be available for download.

\appendix

\section{Ethics}
This work does not raise any ethical issues.

\bibliographystyle{acm}
\bibliography{bibliography}

\end{document}